\documentclass[submission,copyright,creativecommons]{eptcs}

\providecommand{\event}{MeTRiD 2018}

\usepackage{underscore}
\usepackage{subfigure}
\usepackage{mathtools}
\usepackage{verbatim}
\usepackage{stmaryrd}
\usepackage{amsmath}
\usepackage{amssymb}
\usepackage{amsthm}
\usepackage{array}
\usepackage{bm}
\usepackage[capitalize,noabbrev]{cleveref}
\usepackage[colorinlistoftodos,textsize=small]{todonotes}

\theoremstyle{plain}

\theoremstyle{definition}
\newtheorem{definition}{Definition}[section]
\newtheorem{example}{Example}[section]

\newcommand{\List}{\operatorname{lst}}
\newcommand{\sch}[2]{\ensuremath{{\bf #1}\llbracket #2 \rrbracket}}

\usepackage[scaled=0.8]{DejaVuSansMono}
\usepackage{listings}
\newcommand\xqed[1]{\leavevmode\unskip\penalty9999 \hbox{}\nobreak\hfill\quad\hbox{#1}}

\newcommand\EOE{\xqed{$\triangle$}}

\newcommand{\supp}{\operatorname{supp}}
\newcommand{\node}{\operatorname{node}}
\newcommand{\len}{\operatorname{len}}

\newcommand{\flatten}{\operatorname{flatten}}

\newcommand{\wfc}{\operatorname{surg}}
\newcommand{\ode}{\mathrm{ODE}}
\newcommand{\csp}{\mathrm{CSP}}
\newcommand{\io}{\mathrm{IO}}

\newcommand{\calC}{\mathcal{C}}
\newcommand{\calD}{\mathcal{D}}

\newcommand{\calN}{\mathcal{N}}

\newcommand{\calP}{\mathcal{P}}

\newcommand{\calV}{\mathcal{V}}

\newcommand{\NN}{\mathbb{N}}

\newcommand{\RR}{\mathbb{R}}

\newcommand{\dom}{\operatorname{dom}}

\title{Treo: Textual Syntax for Reo Connectors}
\author{Kasper Dokter \qquad Farhad Arbab
\institute{Formal methods\\Centrum Wiskunde \& Informatica\\Amsterdam, Netherlands}
\email{\{K.P.C.Dokter,Farhad.Arbab\}@cwi.nl}
}
\def\titlerunning{Treo: Textual Syntax for Reo Connectors}
\def\authorrunning{Dokter, Arbab}

\begin{document}

\pagestyle{headings} 

\maketitle

\begin{abstract}
Reo is an interaction-centric model of concurrency for compositional specification of communication and coordination protocols.  Formal verification tools exist to ensure correctness and compliance of protocols specified in Reo, which can readily be (re)used in different applications, or composed into more complex protocols. Recent benchmarks show that compiling such high-level Reo specifications produces executable code that can compete with or even beat the performance of hand-crafted programs written in languages such as C or Java using conventional concurrency constructs.

The original declarative graphical syntax of Reo does not support intuitive constructs for parameter passing, iteration, recursion, or conditional specification. 
This shortcoming hinders Reo's uptake in large-scale practical applications.  
Although a number of Reo-inspired syntax alternatives have appeared in the past, none of them follows the primary design principles of Reo: 
a) declarative specification; b) all channel types and their sorts are user-defined; and c) channels compose via shared nodes. 
In this paper, we offer a textual syntax for Reo that respects these principles and supports flexible parameter passing, iteration, recursion, and conditional specification.
In on-going work, we use this textual syntax to compile Reo into target languages such as Java, Promela, and Maude.
\end{abstract}

\section{Introduction}
\label{sec:intro}

The advent of multicore processors has intensified the significance of coordination in concurrent applications. 
A programmer tackles the coordination concern of an application by specifying a (usually implicit) protocol 
that defines all possible permissible interactions among different active components of the application. 
Depending on the language used, programmers define their protocols at different levels of abstraction.
A threading library, for instance, generally offers only basic synchronization primitives, such as locks and semaphores, 
that can be inserted into imperative code to ensure execution follows an implicitly defined protocol.
Exogenous coordination languages offer syntax to programmers to explicitly define their interaction protocols at a high level of abstraction. 

Reo \cite{DBLP:journals/mscs/Arbab04, DBLP:conf/birthday/Arbab11} is an example of such a coordination language that defines an interaction protocol as a {\em connector}: 
a graph-like structure that enables (a)synchronous data flow along its edges (cf., \cref{fig:alternator}).
Each edge, called a {\em channel}, has a user-defined type and two channel ends.
The type determines the behavior of the channel, specified as a constraint on the flows of data at its two ends.
This constraint is expressed in a used-defined semantic sort, such as timed data streams, constraint automata, or coloring semantics \cite{DBLP:journals/cuza/JongmansA12}.
A channel end is either a {\em source end} through which the channel accepts data, or a {\em sink end} through which the channel offers data.
Multiple channel ends coincident at a vertex of the connector together form a {\em node}.
Nodes have predefined `merge-replicate' behavior: 
a node repeatedly accepts a datum from one of its coincident sink ends, chosen non-deterministically, 
and offers a copy of that datum through every one of its coincident source ends. 

Tools for Reo have been implemented as a collection of Eclipse plugins called the ECT \cite{ECT}.
The main plugin in this tool set consists of a graphical editor that allows a user to draw a connector on a canvas. 
The graphical editor has an intuitive interface with a flat learning curve.
However, it does not provide constructs to express parameter passing, iteration, recursion, or conditional construction of connector graphs. 
Such language constructs are more easily offered by familiar programming language constructs in a textual representation of connectors.

In the context of Vereofy (a model checker for Reo), Baier, Blechmann, Klein, and Kl\"uppelholz developed the Reo Scripting Language (RSL) 
and its companion language, the Constraint Automata Reactive Module Language (CARML) \cite{DBLP:conf/coordination/BaierBKK09,DBLP:books/daglib/0031251}.
RSL is the first textual language for Reo that includes a construct for iteration, and a limited form of parameter passing. 
Primitive channels and nodes are defined in CARML, a guarded command language for specification of constraint automata. 
Programmers then combine CARML specified constraint automata as primitives in RSL to construct complex connectors and/or complete systems.
In contrast to the declarative nature of the graphical syntax of Reo, RSL is imperative.

Jongmans developed the First-Order Constraint Automata with Memory Language (FOCAML) \cite{Jongmans16}, 
a textual declarative language that enables compositional construction of connectors from a (pre-defined set of) primitive components.
As a textual representation for Reo, however, FOCAML has poor support for its primary design principle: 
Reo channels are user-defined, not tied to any specific formalism to express its semantics, and compose via shared nodes with predefined merge-replicate behavior.
Although FOCAML components are user-defined, FOCAML requires them to be of the same predefined semantic sort (i.e., constraint automata with memory \cite{DBLP:journals/scp/BaierSAR06}).
The primary concept of Reo nodes does not exist in FOCAML, which forces explicit construction of their `merge-replicate' behavior in FOCAML specifications.

Jongmans et al. have shown by benchmarks that compiling Reo specifications
can produce executable code whose performance competes with or even beats that of hand-crafted programs written in
languages such as C or Java using conventional concurrency constructs \cite{DBLP:conf/coordination/JongmansHA14,DBLP:conf/coordination/JongmansA15,DBLP:conf/tacas/JongmansA16,DBLP:journals/corr/JongmansA16,DBLP:journals/scp/JongmansA18}. 
A textual syntax for Reo that preserves its declarative, compositional  nature, allows user-defined primitives, 
and faithfully complies with the semantics of its nodes can significantly facilitate the uptake of Reo for specification of protocols in large-scale practical applications. 

In this paper, we introduce Treo, a declarative textual language for component-based specification of Reo connectors 
with user-defined semantic sorts and predefined node behavior.
We recall the basics of Reo (\cref{sec:reo}).
We describe the structure of a Treo file by means of an abstract syntax (\cref{sec:syntax}).
In Listing \ref{lst:concretesyntax}, we provide a concrete syntax of Treo as an ANTLR4 grammar \cite{Parr13}.
In on-going work, we currently use Treo to compile Reo into target languages such as Java, Promela, and Maude \cite{GitHub}.
The construction of the Treo compiler is based on the theory of stream constraints \cite{DA18}. 

In order to preserve the agnosticism of Reo regarding the concrete semantics of its primitives, 
Treo uses the notion of user-defined {\em semantic sorts}.
A user-defined semantic sort consist of a set of component instances together with a composition operator $\wedge$, a substitution operator $[\,/\,]$, and a trivial component $\top$ (\cref{sec:types}). 
The composition operator defines the behavior of composite components as a composition of its operands.
The substitution operator binds nodes in the interface or passes values to parameters.

For a given semantic sort, we define the meaning of abstract Treo programs (\cref{sec:semantics}).
Treo is very liberal with respect to parameter values.
A component definition not only accepts the usual (structured) data as actual parameters, but also other component instances and other component definitions. 
Among other benefits, this flexible parameter passing supports {\em component sharing}, which is useful to preserve component encapsulation \cite[Figure 2]{DBLP:journals/spe/BrunetonCLQS06}.

A given semantics sort may possibly distinguish between inputs and outputs.
Thus, not all combinations of components may result in a valid composite component.
For example, the composition may not be defined, if two components share an output.
In Treo, however, it is safe to compose components on their outputs, 
because, complying with the semantics of Reo, the compiler inserts special {\em node components} to ensure well-formed compositions (\cref{sec:direction}).

We conclude by discussing related work (\cref{sec:related}), and pointing out future work (\cref{sec:conclusion}).

\section{Reo}
\label{sec:reo}

We briefly recall the basics of the Reo language and refer to \cite{DBLP:journals/mscs/Arbab04} and \cite{DBLP:conf/birthday/Arbab11} for further details.
Reo is a language for specification of interaction protocols, originally proposed with a graphical syntax.
A Reo program, called a {\em connector}, is a graph-like structure whose edges consist of {\em channels} that enable synchronous and asynchronous data flow and whose vertices consist of {\em nodes} that synchronously route data among multiple channels.
Each channel has a type and two channel ends.
Each channel end is either a {\em source end}, through which the channel accepts data, or a {\em sink end}, through which the channel offers data.
The type of a channel completely defines the behavior of the channel in some user-defined semantic sort.  
Reo is agnostic regarding the semantic sort that expresses the behavior of its channel types, 
so long as the semantic sort preserves Reo's compositional construction principle (cf., \cref{defn:types}).
\cref{tab:channelsnodes} shows some frequently used channels and an example node together with an informal description of their behavior.

\begin{table}[ht]
    \center
    \renewcommand{\arraystretch}{1.5}
    \begin{tabular}[t]{>{\centering\arraybackslash}m{3.5cm} >{\centering\arraybackslash}m{11cm}}
        \includegraphics[]{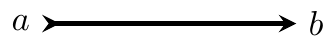} & A {\sf Sync} channel accepts datum from its source end $a$, when its simultaneous offer of this datum at its sink end $b$ succeeds. \\ \hline
        \includegraphics[]{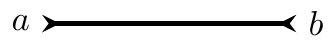} & A {\sf SyncDrain} channel simultaneously accepts data from both its source ends $a$ and $b$ and loses the data. \\ \hline
        \includegraphics[]{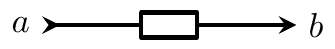} & An empty {\sf FIFO$_1$} accepts data from its source end $a$ and becomes a full {\sf FIFO$_1$}. 
        A full {\sf FIFO$_1$} offers its stored data at its sink end $b$ and, when its offer succeeds, it becomes an empty {\sf FIFO$_1$} again. \\ \hline
        \includegraphics[]{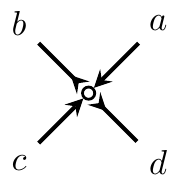} & A Reo node accepts a datum from one of its coincident sink ends ($a$ or $c$), when its simultaneous offer to dispense a copy of this datum through every one of its coincident source ends ($b$ and $d$) succeeds.
    \end{tabular}
    \caption{Informal description of the behavior of nodes and of some channels in Reo.}
    \label{tab:channelsnodes}
\end{table}

\begin{figure}[ht]
    \centering
    \includegraphics[scale=1]{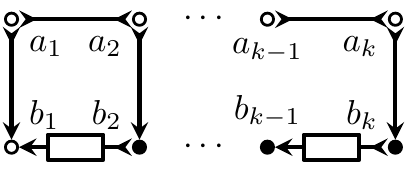}
    \caption{Construction of the {\sf Alternator$_k$} Reo connector, for $k \geq 2$.}
    \label{fig:alternator}
\end{figure}

The key concept in Reo is composition, which allows a programmer to build complex connectors out of simpler ones.
For example, using the channels in \cref{tab:channelsnodes}, we can construct the {\sf Alternator$_k$} connector, for $k \geq 2$, as shown in \cref{fig:alternator}.
For $k=2$, the {\sf Alternator$_2$} consists of four nodes ($a_1$, $a_2$, $b_1$, and $b_2$) and 
four channels, namely a {\sf SyncDrain} channel (between $a_1$ and $a_2$), two {\sf Sync} channels (from $a_1$ to $b_1$, and from $a_2$ to $b_2$), and a {\sf FIFO$_1$} channel (from $b_2$ to $b_1$).

The behavior of the {\sf Alternator$_2$} connector is as follows. 
Suppose that the environment is ready to offer a datum at each of the nodes $a_1$ and $a_2$, and ready to accept a datum from node $b_1$.
According to \cref{tab:channelsnodes}, nodes $a_1$ and $a_2$ both offer a copy of their received datum to the {\sf SyncDrain} channel.
The {\sf SyncDrain} channel ensures that nodes $a_1$ and $a_2$ accept data from the environment only simultaneously.
The {\sf Sync} channel from $a_1$ to $b_1$ ensures that node $b_1$ simultaneously obtains a copy of the datum offered at $a_1$.
By definition, node $b_1$ either accepts a datum from the connected {\sf Sync} channel or it accepts a datum from the {\sf FIFO$_1$} channel (but not from both simultaneously), 
and offers this datum immediately to the environment.
Because the {\sf FIFO$_1$} is initially empty, $b_1$ has no choice but to accept and dispense the datum from $a_1$. 
Simultaneously, the {\sf Sync} channel from $a_2$ to $b_2$ ensures that the value offered at $a_2$ is stored in the {\sf FIFO$_1$} buffer.
In the next step, the environment at node $b_1$ has no choice but to retrieve the datum in the buffer, after which the behavior repeats.

\section{Treo syntax}
\label{sec:syntax}

We now present a textual representation for the graphical Reo connectors in \cref{sec:reo}.
\cref{tab:syntax} shows the abstract syntax of Treo. 

\begin{table}[th]
    \centering
    \begin{tabular}{ll}
    $K \ ::= \ I \mid KND$                         & $D \ ::= \ V \mid \langle U_0 \rangle (U_1) \{ C \} $ \\
    $L \ ::= \ \epsilon \mid L,T \mid L,T_0..T_1 $ & $C \ ::= \ V \mid A \mid C_0 C_1 \mid \{ C \bm{\mid} P \} \mid D \langle L \rangle (U)$ \\ 
    $U \ ::= \  \epsilon \mid U, V$                & $T \ ::= \ V \mid C \mid D \mid [L] \mid T_0:T_1 \mid T[L] \mid F(L)$ \\
    $V \ ::= \ N \mid V[L]$                        & $P \ ::= \ V \in T \mid R(L) \mid \neg P \mid P_0 \wedge P_1 \mid P_0 \vee P_1 \mid (P)$ \\  
    \end{tabular}
    \caption{Abstract syntax of Treo, with start symbol $K$ (a source file), and terminal symbols for imports ($I$), primitive components ($A$), functions ($F$), relations ($R$), names ($N$), and the empty list ($\epsilon$). The bold vertical bar in $\{C \bm{\mid} P\}$ is just text. }
    \label{tab:syntax}
\end{table}

We introduce the symbols in the abstract syntax by identifying them in some concrete examples.
These concrete examples are Treo programs that can be parsed using the concrete Treo syntax shown in Listing \ref{lst:concretesyntax}.

\begin{lstlisting}[caption={Concrete ANTLR4 syntax of Treo ({\sf Treo.g4}).}, 
label={lst:concretesyntax}]
grammar Treo;
file    : sec? imp* assg* EOF;
sec     : 'section' name ';' ;
imp     : 'import' name ';' ;
assg    : ID defn ;
defn    : var | params? nodes comp ;
comp    : defn vals? args | var | '{' atom+ '}' | '{' comp* ('|' pred)? '}' 
        | 'for' '(' ID 'in' list ')' comp 
        | 'if' '(' pred ')' comp ('else' '(' pred ')' comp)* ('else' comp)? ;
atom    : STRING ; /* Example syntax for primitive components */
pred    : 'true' | 'false' | '(' pred ')' | var 'in' list 
        | term op=('<=' | '<' | '>=' | '>' | '=' | '!=') term 
        | var | 'forall' ID 'in' list ':' pred | 'exists' ID 'in' list ':' pred
        | 'not' pred | pred ('and'|',') pred | pred 'or' pred | pred 'implies' pred ;
term    : var | NAT | BOOL | STRING | DEC | comp | defn | list | 'len(' term ')'
        | '(' term ')' | <assoc=right> term list | <assoc=right> term '^' term 
        | '-' term | term op=('*' | '/' | '%' | '+' | '-') term ;
vals    : '<' '>' | '<' term (',' term)* '>' ;
list    : '[' ']' | '[' item (',' item)* ']' ;
item    : term | term '..' term | term ':' term ;
args    : '(' ')' | '(' var (',' var)* ')' ; 
params  : '<' '>' | '<' var (',' var)* '>' ; 
nodes   : '(' ')' | '(' node (',' node)* ')' ;
node    : var (io=('?' | '!' | ':') ID?)? ;
var     : name list* ;
name    : (ID '.')* ID ;
NAT     : ('0' | [1-9][0-9]*) ;
DEC     : ('0' | [1-9][0-9]*) '.' [0-9]+ ;
BOOL    : 'true' | 'false' ;
ID      : [a-zA-Z_][a-zA-Z0-9_]*;
STRING  : '\"' .*? '\"' ;
SPACES  : [ \t\r\n]+ -> skip ;
SL_COMM : '//' .*? ('\n'|EOF) -> skip ;
ML_COMM : '/*' .*? '*/' -> skip ;
\end{lstlisting}

Consider the following Treo file ($K$ in \cref{tab:syntax}) representing the {\sf Alternator$_2$}:
\begin{verbatim}
import syncdrain; import sync; import fifo1;
alternator2(a1,a2,b1) { sync(a1,b1) syncdrain(a1,a2) sync(a2,b2) fifo1(b2,b1) }
\end{verbatim}
On the first line, we import ($I$) three different {\em component definitions}.
On the second line, we define the {\tt alternator2} component ($ND$).
Its definition ($D$) has no parameters ($\langle U_0 \rangle$), and three nodes, {\tt a1}, {\tt a2}, and {\tt b1}, in its interface ($(U_1)$).
The body ($\{C\}$) of this definition consists of a set of {\em component instances} that interact via shared nodes.
The first component instance {\tt sync(a1,b1)} is an instantiation ($D \langle L \rangle (U)$) of the imported {\tt sync} definition ($D$) with nodes {\tt a1} and {\tt b1} ($(U)$) and without any parameters ($\langle L \rangle$).

All nodes that occur in the body, but not in the interface, are hidden. 
Hiding renames a node to a fresh inaccessible name, which prevents it from being shared with other components. 
In the case of {\tt alternator2}, node {\tt b2} is not part of the interface, and hence hidden.

Constructed from existing components, {\tt alternator2} is a {\em composite} component $(C_0C_1$).
However, not every component is constructed from existing components, and we call such components {\em primitive} ($A$).
The following Treo code shows a possible (primitive) definition of the {\tt fifo1} component.
\begin{verbatim}
fifo1(a?,b!) { empty -{a},true-> full; full -{b},true-> empty; }
\end{verbatim}
The definition of the {\tt fifo1} differs from the definition of the {\tt alternator2} in two ways.

The first difference is that the {\tt fifo1} component is (in this case) defined directly as a {\em constraint automaton} \cite{DBLP:journals/scp/BaierSAR06}.
Constraint automata constitute a popular semantic sort for specification of Reo component types, and forms the basis of the Lykos compiler \cite{Jongmans16}.
However, constraint automata are not the {\em de facto} standard:
the literature offers more than thirty different semantic sorts for specification of Reo components \cite{DBLP:journals/cuza/JongmansA12}, such as the coloring semantics and timed data stream semantics.
To accommodate the generality that disparate semantics allow, Treo features {\em user-defined semantic sorts}, which means that the syntax for primitive components is user-defined.
For example, this means that we may also define the {\tt fifo1} component by referring to a Java file via \verb|fifo1(a?,b!){ "MyFIFO1.java" }|.

The second difference is that the nodes {\tt a} and {\tt b} in the interface are {\em directed}.
That is, each of its interface nodes is either of type input or output, designated by the markers {\tt ?} and {\tt !}, respectively.
In Reo, it is safe to join two channels on a shared sink node (e.g., node $b_1$ in \cref{fig:alternator}).
However, the composition operators in most Reo semantics do not automatically produce the correct behavior for such nodes (e.g., see \cite[Section 4.3]{DBLP:journals/scp/BaierSAR06} for further details).
Therefore, most Reo semantics require {\em well-formed} compositions, wherein each node has at most one input channel end and at most one output channel end.

The restriction of well-formed compositions can be very inconvenient in practice.
To ensure well-formed compositions, a programmer must implement every Reo node with more than one input or output channel end as a {\em node component}.
The interface of this node component is determined by its {\em degree}, which is a pair $(i,o)$ giving the numbers of its coincident source and sink ends. 
Such explicit node components make component constructions verbose and hard to maintain.
For convenience, the Treo compiler uses the above input/output annotations to compute the degree of each node in a composition, and subsequently inserts the correct node components in the construction.
We may view the input/output annotations as syntactic sugar that ensures well-formed compositions.
This feature allows programmers to remain oblivious to these annotations and well-formed composition.

The ellipses in \cref{fig:alternator} signify the parametrized construction of the {\sf Alternator$_k$} connector, for $k > 2$.
This notation is informal and not supported in the graphical Reo editor \cite{ECT}, which offers no support for parametrized constructions. 
In Treo, however, we can define the {\sf Alternator$_k$} connector as:
\begin{verbatim}
alternator<k>(a[1:k],b[1]) { sync(a[1],b[1]) 
    { syncdrain(a[i-1],a[i]) sync(a[i],b[i]) fifo1(b[i],b[i-1]) | i in [2..k] } }
\end{verbatim}
The definition of the {\tt alternator} depends on a parameter {\tt k}. 
Since Treo is a strongly typed language with type-inferencing, there is no need to specify a type for the (integer) parameter {\tt k}.
The interface consists of an array of nodes {\tt a[1:k]} and the single node {\tt b[1]}.
Here, {\tt [1:k]} is an abbreviation for the list {\tt [[1..k]]} that contains a single list of length {\tt k}.
The array {\tt a[1:2]} stands for the {\em slice} {\tt [a[1],a[2]]} of {\tt a}, while the expression {\tt a[1..2]} stands for the element {\tt a[1][2]} in  {\tt a} (cf., \cref{eq:access}).
For iteration, we write \verb+{ ... | i in [2..k] }+ using set-comprehension ($\{ C \bm{\mid} P \}$ in \cref{tab:syntax}).

Instead of defining {\tt alternator} iteratively, we may also provide a recursive definition as follows:
\begin{verbatim}
recursive_alternator(a[1:k],b[1],b[k]) { recursive_alternator(a[1:k-1],b[1],b[k-1]) 
    { syncdrain(a[k-1],a[k]) sync(a[k],b[k]) fifo1(b[k-1],b[k]) | k > 1 } }
\end{verbatim}
Here, the value of {\tt k} is defined by the size of {\tt a[1:k]}, and we use set-comprehension \verb+{ ... | k > 1 }+ for conditional construction, as well.
Indeed, the resulting set of component instances is non-empty, only if {\tt k  >  1} holds.
Although Treo syntax allows recursive definitions, the semantics presented in \cref{sec:semantics} does not yet support recursion, which we leave as future work.

We illustrate the practicality of Treo by providing code for a chess playing program \cite[Figure 3.29]{Jongmans16}.
In this program, two teams of chess engines compete in a game of chess. We define a chess team as the following Treo component:
\begin{verbatim}
import parse; /* and the other imports */
team<engine[1:n]>(inp,out) { 
    for (i in [1..n]) { 
        engine[i](inp,best[i]) parse(best[i],p[i])
        if (i > 1) concatenate(a[i-1],p[i],a[i]) }
    sync(best[1],a[1]) majority(a[n],b) syncdrain(b,c) 
    fifo1(inp,c) move(b,d) concatenate(c,d,out) }
\end{verbatim}
The for-loop \verb|for (i in [1..n]) ...| and if-statement \verb|if (i > 1) ...| are just syntactic sugar for set-comprehensions \verb+{ ... | i in [1..n] }+ and \verb+{ ... | i > 1 }+, respectively.
The {\tt team} component depends on an array {\tt engine[1:n]} of parameters. 
This array does not contain the usual data values, but consists of Treo component definitions.
In the body of the {\tt team} component, these definitions are instantiated via {\tt engine[i](inp,best[i])}.
In RSL \cite{DBLP:conf/coordination/BaierBKK09,DBLP:books/daglib/0031251} and FOCAML \cite{Jongmans16}, it is impossible to pass a component as a parameter, which makes these languages less expressive than Treo.

We may view the {\tt team} component as an example of role-oriented programming \cite{DBLP:conf/birthday/ChrszonDB0K16}.
Indeed, the {\tt team} component encapsulates a list of chess engines in a component, so that they can collectively be used as a single participant in a chess match:
\begin{verbatim}
match() { fifo1full<"">(a,b) fifo1(c,d) team<[eng1, eng2]>(a,d) team<[eng3]>(b,c) }
\end{verbatim}
Treo treats not only component definitions, but also component instances as values.
By passing a single component instance as a parameter to multiple components, this feature allows {\em component (instance) sharing} (cf., \cite[Figure 2]{DBLP:journals/spe/BrunetonCLQS06}).
Hence, it is straightforward to implement a chess match, wherein a single instance of a chess engine plays against itself.

\section{Semantic sorts}
\label{sec:types}

As noted in \cref{sec:syntax}, Reo channels can be defined in many different semantic formalisms \cite{DBLP:journals/cuza/JongmansA12}, such as the constraint automaton semantics, the coloring semantics, or the timed data stream semantics.
Although each sort of Reo semantics has its unique properties, each of them can be used to define a collection of composable components with parameters and nodes, which we call a {\em semantic sort}:

\begin{definition}[Semantic sort]
    \label{defn:types}
    A semantic sort over a set of names $\calN$ with values from $\calV$ is a tuple $(\calC, \wedge, [\, /\, ], \top)$ that consists of 
    a set of components $\calC$, 
    a composition operator $\wedge : \calC \times \calC \longrightarrow \calC$, 
    a substitution operator $[\, /\, ] : \calC \times (\calN \cup \calV) \times \calN \longrightarrow \calC$, and 
    a trivial component $\top \in \calC$.
\end{definition}

We assume that the set of names and the set of values are disjoint, i.e., $\calN \cap \calV = \emptyset$.
For convenience, we write $C \wedge C'$ for $\wedge(C,C')$, and $C[y/x]$ for $[\, /\, ](C,y,x)$.
For any semantic sort $T$, we write $\calC_T$ for its set of components, $\wedge_T$ for its composition operator, $[\, / \, ]_T$ for its substitution operator, and $\top_T$ for its trivial component.
The composition operator $\wedge_T$ ensures that the behavior of finite non-empty compositions is well-defined.
To empty compositions we assign the trivial component $\top_T$.
The substitution operator $[\,/\,]_T$ allows us to change the interface of a component via renaming or instantiation.
Let $C \in \calC_T$ be a component and $x \in \calN$ a name.
For a name $y \in \calN$, the construct $C[y/x]_T$ renames every occurrence of name $x$ in $C$ to $y$.
For a value $y \in \calV$, the construct $C[y/x]_T$ instantiates (parameter) $x$ in $C$ to $y$.
(See \cref{ex:io} for an example of the distinction between renaming and instantiations.)

A semantic sort $T$ implicitly defines an interface for each component $C \in \calC$ via the map $\supp : \calC_T \longrightarrow 2^\calN$ defined as $\supp(C) \ = \ \{ x \in \calN \mid C[y/x]_T \neq C \text{, for some name } y \in \calN \}$.
If name $x$ does not `occur' in $C$, substitution of $x$ by any name $y$ does not affect $C$, i.e., $C[y/x]_T = C$.

\begin{example}[Systems of differential equations]
    \label{ex:ode}
    The set $\ode$ of systems of ordinary differential equations with variables from $\calN$ and values $\calV = \{v : \RR \longrightarrow \RR\}$ constitute a semantic sort. Composition is union, substitution is binding a name or value to a given name, and the trivial component is the empty system of equations.
    Using the $\ode$ semantic sort, we can define continuous systems in Treo.
    \EOE
\end{example}

\begin{example}[Process calculi]
    \label{ex:csp}
    Consider the process calculus CSP, proposed by Hoare \cite{DBLP:journals/cacm/Hoare78}.
    The set $\csp$ of all such process algebraic terms comprises a semantic sort. 
    Each process can participate in a number of events, which we can interpret as names from a given set $\calN$. 
    We model the composition of CSP processes $P$ and $Q$ by means of the interface parallel operator $P \ |[X]| \ Q$, where $X \subseteq \calN$ is the set of event names shared by $P$ and $Q$. 
    We define substitution as simply (1) renaming the event, if a name is substituted for an event; or (2) hiding the event, if a values is substituted for an event.
    Since neither STOP nor SKIP shares any event with its environment, we may use either one to denote the trivial component.
    \EOE
\end{example}

\begin{example}[I/O-components]
    \label{ex:io}
    Let $T$ be a semantic sort over $\calN$ and $\calV$.
    We define the I/O-component sort $\io_T$ over $T$ using the notion of a primitive I/O-component of sort $T$. 
    
    A {\em primitive I/O-component} $P$ of sort $T$ is a tuple $(C, I, O)$, where 
    $C \in \calC_T$ is a component, 
    $I \subseteq \calN$ is a set of input names, 
    $O \subseteq \calN$ is a set of output names.
    For $P \subseteq \calN$ and $x \in \calN$ and $y \in \calN \cup \calV$, define
    \begin{equation}
        P[y/x] \ = \ 
        \begin{cases}
            (P-\{x\}) \cup \{y\}     & \text{if } x \in P \text{ and } y \in \calN \\
            P-\{x\}                  & \text{if } x \in P \text{ and } y \in \calV \\
            P                        & \text{otherwise}
        \end{cases}
    \end{equation}   
    We define substitution on primitive I/O-components as $(C, I, O)[y/x] = (C[y/x], I[y/x], O[y/x])$, for all $x \in \calN$ and $y \in \calN \cup \calV$.
    We denote the set of primitive I/O-components over $T$ as $\calP_T$.
    
    An {\em I/O-component} of sort $T$ is a sequence $P_1 \cdots P_n \in \calP_T^\ast$, with $n \geq 0$, of primitive I/O-components of sort $T$.
    Composition of I/O-components is concatenation $\cdot$ of sequences.
    The trivial I/O-component is the empty sequence $\epsilon$. 
    We define substitution of composite I/O-components as $(P_1 \cdots P_n)[y/x] = P_1[y/x] \cdots P_n[y/x]$, for all $x \in \calN$ and $y \in  \calN \cup \calV$.
    Hence, $\io_T = (\calP_T^\ast, \cdot, [\,/\,], \epsilon)$ is a semantic sort.
    \EOE
\end{example}

\section{Denotational semantics}
\label{sec:semantics}

We define the denotational semantics of the Treo language over a fixed, but arbitrary, semantic sort $T$.
The main purpose of this denotational semantics is to provide a clear abstract structure that guides the implementation of Treo parsers.
The syntax to which this denotational semantics applies is the abstract syntax in \cref{tab:syntax}.
The general structure of our denotational semantics is quite standard, and adheres to Schmidt's notation \cite{Schmidt97}.

Although Treo syntax allows recursive definitions, the semantics presented in this section does not support this feature. 
Since not all recursive definitions define finite compositions of components, extending the current semantics with recursion is not straightforward, and we leave it as future work. 

Variables and terms in Treo are structured as non-rectangular arrays.
The set of all {\em (ragged) arrays} over a set $X$ is the smallest set $X^\square$ such that both $X \subseteq X^\square$ and $[x_0,\ldots,x_{n-1}] \in X^\square$, if $n\geq 0$ and $x_i \in X^\square$ for all $0 \leq i < n$. 
For example, the set $\NN^\square$ of ragged arrays over integers contains all natural numbers from $\NN$ as `atomic' arrays, as well as the array $[37, [], [[2, [55], 3]]] \in \NN^\square$.
Every ragged array has a length, which can be computed via the map $\len : X^\square \longrightarrow \NN$ defined inductively as $\len(x) = 0$, if $x \in X$, and $\len([x_0,\ldots,x_{n-1}]) = n$, otherwise.
If $x = [x_0,\ldots,x_{n-1}] \in X^\square$ is a ragged array, we access its entries via the function application $x(i) = x_i$, for every $0 \leq i < n$.
We extend the access map $\NN^\square$ by defining
\begin{equation}
\label{eq:access}
  x([i_0,\ldots,i_n]) = 
\begin{cases}
    x(i_0)([i_1,\ldots,i_n]) & \text{if } i_0 \in \NN \\
    [x(i_{00})([i_1,\ldots,i_n]), \ldots, x(i_{0m})([i_1,\ldots,i_n])] & \text{if } i_0 = [i_{00},\ldots,i_{0m}]
\end{cases},
\end{equation}
whenever the right-hand side is defined.
Two ragged arrays $x \in X^\square$ and $y \in Y^\square$ have the same structure ($x \simeq y$) iff $x \in X$ and $y \in Y$, or $\len(x) = \len(y)$ and $x(i) \simeq y(i)$ for all $0 \leq i < \len(x)$.
We can flatten a ragged array from $X^\square$ to a sequence over $X$ via the map $\flatten : X^\square \longrightarrow X^\ast$ defined as $\flatten(x) = x$, if $x \in X$, and $\flatten([x_0,\ldots,x_{n-1}]) = \flatten(x_0) \cdots \flatten(x_{n-1})$, otherwise.

Suppose that semantic sort $T$ is defined over a set of names $\calN$ and a set of values $\calV$, with $\calN \cap \calV = \emptyset$.
For simplicity, we assume that, for every component $C \in \calC_T$, its support $\supp(C) \subseteq \calN$ is finite.
Since Treo views components as values, we assume the inclusion $\calC_T \subseteq \calV$.

We assume that the set of names $\calN$ is closed under taking subscripts from $\NN$.
That is, if $x \in \calN$ is a name and $i \in \NN$ is a natural number, then we can construct a fresh name $x_i \in \calN$.
To construct sequences of data with variable lengths, we use a map 
$\List : \NN^2 \longrightarrow \NN^\square$ 
that constructs from a pair $(i,j) \in \NN^2$ of integers a finite ordered list $[i, i+1, \ldots, j]$ in $\NN^\square$.

Recall from \cref{sec:syntax} that a component accepts an arbitrary but finite number of parameters and nodes. 
Therefore, we define a component definition as a map $D : \calV^\square \times \calN^\square \longrightarrow \calC_T \cup \{ \lightning \}$
that takes an array of parameter values from $\calV^\square$ and an array of nodes from $\calN^\square$ and returns a component or an {\em error} $\lightning$.
Let $\calD = (\calC_T \cup \{ \lightning \})^{\calV^\square \times \calN^\square}$ be the set of all definitions.
As mentioned earlier, Treo also allows definitions as values, which amounts to the inclusion $\calD \subseteq \calV$.\footnote{\label{foot:values}
Such a set of values $\calV$ exists only if $\calV \mapsto \calC_T \cup  (\calC_T \cup \{ \lightning \})^{\calV^\square \times \calN^\square}$ admits a pre-fixed point. In this work, we simply assume that such $\calV$ exists.}

We evaluate every Treo construct in its {\em scope} $\sigma : N \longrightarrow \calV^\square$, with $N \subseteq \calN$ finite, which assigns a value to a finite collection of locally defined names. 
We write $\Sigma = \{ \sigma : N \longrightarrow \calV^\square \mid N \subseteq \calN \mbox{ finite} \}$ for the set of scopes.
For a name $x \in \calN$ and a value $d \in \calV^\square$, we have a scope $\{x \mapsto d\} :  \{x\} \longrightarrow \calV^\square$ defined as $\{x \mapsto d\}(x) = d$.
For any two scopes $\sigma,\sigma' \in \Sigma$, we have a composition $\sigma\sigma' \in \Sigma$ such that for every $x \in \dom(\sigma) \cup \dom(\sigma')$ we have $(\sigma\sigma')(x) = \sigma'(x)$, if $x \in \dom(\sigma')$, and $(\sigma\sigma')(x) = \sigma(x)$, otherwise.
The composite scope $\sigma\sigma'$ can be viewed as an extension of $\sigma$ that includes definitions and updates from $\sigma'$.

Let Names be the set of parse trees with root $N$, and let $\sch{N}{-} : {\rm Names} \longrightarrow \calN$ be the semantics of names. 
We define the semantics of variables as a map $\sch{V}{-} : {\rm Variables} \longrightarrow (\calN^\square \cup \{\lightning\})^\Sigma,$
where Variables is the set of parse trees with root $V$.
For a scope $\sigma \in \Sigma$, we define $\sch{V}{-}(\sigma)$ as follows:
\begin{enumerate}
    \item $\sch{V}{N}(\sigma) = \sch{N}{N}$;
    
    \item $\sch{V}{V[L]}(\sigma) = \begin{cases}
x(k) & \text{if } \sch{V}{V}(\sigma) = x \in \calN^\square \text{ and } \sch{L}{L}(\sigma) = k \in \NN^\square \\
\lightning & \text{otherwise}
    \end{cases}.$
\end{enumerate}
Since $\calN$ is closed under taking subscripts, we can define $n(i) = n_i$, for all $n \in \calN$ and $i \in \NN$, which ensures that $x(k) \in \calN^\square$ is always defined.

The semantics of arguments is a map $\sch{U}{-} : {\rm Arguments} \longrightarrow (\calN^\square \cup \{\lightning\})^\Sigma,$
where Arguments is the set of all parse trees with root $U$. 
For a scope $\sigma \in \Sigma$, we define $\sch{U}{-}(\sigma)$ as follows:
\begin{enumerate}
    \item $\sch{U}{\epsilon}(\sigma) = []$;
    
    \item $\sch{U}{U,V}(\sigma) = \begin{cases} 
    [x_1,\ldots,x_{n+1}] & \text{if } \sch{U}{U}(\sigma) = [x_1,\ldots,x_n] \text{ and } \sch{V}{V}(\sigma) = x_{n+1}\\ \lightning & \text{otherwise} \end{cases}$.
    \end{enumerate}

Let Functions be the set of parse trees with root $F$, and let $\sch{F}{-} : {\rm Functions} \longrightarrow \{ \calV^k \longrightarrow \calV \mid k \in \NN\}$ be the semantics of functions.
The semantics of terms is a map $\sch{T}{-} : {\rm Terms} \longrightarrow (\calV^\square \cup \{\lightning\})^\Sigma,$
where Terms is the set of parse trees with root $T$. 
For a scope $\sigma \in \Sigma$, we define $\sch{T}{-}(\sigma)$ inductively as follows: 
\begin{enumerate}
    \item $\sch{T}{V}(\sigma) = \begin{cases}\sigma(\sch{V}{V}(\sigma)) & \text {if defined} \\ \lightning & \text{otherwise} \end{cases}$;
    
    \item $\sch{T}{C}(\sigma) = \sch{C}{C}(\sigma)$, which is well-defined since $\calC_T \subseteq \calV$;
    
    \item $\sch{T}{D}(\sigma) = \sch{D}{D}(\sigma)$, which is well-defined since $\calD \subseteq \calV$;	
    
    \item $\sch{T}{[L]}(\sigma) = \sch{L}{L}(\sigma)$;
    
    \item $\sch{T}{T_0:T_1}(\sigma) = \begin{cases} \List(x_0,x_1-1) & \text {if } \sch{T}{T_i}(\sigma) = x_i \in \NN \text{ for } i \in \{0,1\} \\ \lightning & \text{otherwise} \end{cases}$;
    
    \item $\sch{T}{T[L]}(\sigma) = \begin{cases} x(k) & \text{if } \sch{T}{T}(\sigma) = x \in \calV^\square \text{ and } \sch{L}{L}(\sigma) = k \in \NN^\square\\ \lightning & \text{otherwise}\end{cases}$;
    
    \item $\sch{T}{F(L)}(\sigma) = \begin{cases} \sch{F}{F}(\sch{L}{L}(\sigma)) & \text{if } \sch{F}{F} : \calV^k \longrightarrow \calV \text{ and } \len(\sch{L}{L}(\sigma)) = k\\ \lightning & \text{otherwise} \end{cases}$.
\end{enumerate}

The semantics of lists is a map $\sch{L}{-} : {\rm Lists} \longrightarrow (\calV^\square \cup \{\lightning\})^\Sigma,$
where Lists is the set of parse trees with root $L$. 
For a given scope $\sigma \in \Sigma$, we define $\sch{S}{-}(\sigma)$ inductively as follows:
\begin{enumerate}
    \item $\sch{L}{\epsilon}(\sigma) = []$;
    \item $\sch{L}{L,T}(\sigma) = \begin{cases}
     [x_1,\ldots,x_{n+1}] & \text{if } \sch{L}{L}(\sigma) = [x_1,\ldots,x_n] \in \calV^\square \text{ and } \sch{T}{T}(\sigma) = x_{n+1} \in \calV \\
     \lightning & \text{otherwise}
\end{cases}$;
    \item $\sch{L}{L,T_0..T_1}(\sigma) = \begin{cases} [x_1,\ldots,x_{n+k}] & \text{if } \sch{L}{L}(\sigma) = [x_1,\ldots,x_n] \in \calV^\square, \sch{T}{T_i}(\sigma) = a_i \in \calV, \\ & \text{for } i \in \{0,1\} \text{, and } \List(a_0,a_1) = [x_{n+1},\ldots,x_{n+k}] \\ \lightning & \text{otherwise} \end{cases}$.
\end{enumerate}

Since we use predicates in Treo for list comprehension, we define the semantics of predicates as a map $\sch{P}{-} : {\rm Predicates} \longrightarrow (2^\Sigma)^\Sigma,$ 
where Predicates is the set of all parse trees with root $P$.
For a scope $\sigma \in \Sigma$, we define the semantics $\sch{P}{-}(\sigma)$ of a predicate $P$ as the set of all extensions of $\sigma$ that satisfy $P$.
We define $\sch{P}{-}(\sigma)$ inductively as follows:
\begin{enumerate}	
    \item $\sch{P}{V \in T}(\sigma) = \begin{cases}
    \{\sigma\{x \mapsto t_i\} \mid 1 \leq i \leq n\} & \mbox{if } \sch{V}{V}(\sigma) = x \notin \dom(\sigma), \sch{T}{T}(\sigma) = [t_1,\ldots,t_n] \\
    \{\sigma\}   & \mbox{if } \sch{T}{V}(\sigma) \in \sch{T}{T}(\sigma)  \\
    \emptyset & \mbox{otherwise} 
    \end{cases}$, 
    
    \item If $P$ is $R(L)$, we define $\sch{P}{R(L)}(\sigma) = \{\sigma' \in \Sigma \mid \sigma'\sigma = \sigma', \sch{L}{L}(\sigma') \in \sch{R}{R} \}$;
    
    \item If $P$ is $\neg P$, we define $\sch{P}{\neg P}(\sigma) = \{\sigma' \in \Sigma \mid \sigma'\sigma = \sigma', \neg\sch{P}{P}(\sigma')\}$;
    
    \item If $P$ is $P_0 \wedge P_1$, we define $\sch{P}{P_0 \wedge P_1}(\sigma) = \sch{P}{P_0}(\sigma) \cap \sch{P}{P_1}(\sigma)$;
    
    \item If $P$ is $P_0 \vee P_1$, we define $\sch{P}{P_0 \vee P_1}(\sigma) = \sch{P}{P_0}(\sigma) \cup \sch{P}{P_1}(\sigma)$;
    
    \item If $P$ is $(P)$, we define $\sch{P}{(P)}(\sigma) = \sch{P}{P}(\sigma)$.
\end{enumerate}
For set and list comprehensions, we can iterate over only a finite subset of scopes $\sch{P}{P}(\sigma)$ of $P$.
We ensure this by restricting the set of scopes to those solutions that are minimal with respect to inclusion of domains.
Formally, we write $\min\sch{P}{P}(\sigma)$ for the set of all scopes that are minimal with respect to $\leq$ defined as $\sigma_1 \leq \sigma_2$ iff $\dom(\sigma_1) \subseteq \dom(\sigma_2)$, for all $\sigma_1,\sigma_2 \in \sch{P}{P}(\sigma)$.

The semantics of component instances is a map $\sch{C}{-} : {\rm Components} \longrightarrow (\calC_T \cup \{ \lightning \})^\Sigma,$
where Components is the set of parse trees with root $C$. 
Recall that Treo views components as values ($\calC_T \subseteq \calV$).
Given a scope $\sigma \in \Sigma$, we define $\sch{C}{-}(\sigma)$ inductively as follows:
\begin{enumerate}
    \item $\sch{C}{V}(\sigma) = \begin{cases}
    \sigma(x) & \text{if } \sch{V}{V}(\sigma) = x \in \dom(\sigma) \text{ and } \sigma(x) \in \calC_T \\
 \lightning & \text{otherwise}
\end{cases}$;
    
    \item $\sch{C}{A}(\sigma) = \sch{A}{A}$, where ${\sch{A}{-} : {\rm Atoms} \longrightarrow \calC_T}$ is the semantics of primitive components;
    
    \item $\sch{C}{C_0C_1}(\sigma) = \begin{cases}
    \sch{C}{C_0}(\sigma) \wedge_T \sch{C}{C_1}(\sigma) & \text{if } \sch{C}{C_i}(\sigma) \in \calC_T \text{, for } i \in \{0,1\}\\
    \lightning & \text{otherwise}
\end{cases}$;
    
    \item $\sch{C}{\{ C : P \}}(\sigma) = \begin{cases}
    \top_T & \text{if } \min\sch{P}{P}(\sigma) \text{  is empty or infinite}\\
    C_1 \wedge_T \cdots \wedge_T C_k & \text{if } \min\sch{P}{P}(\sigma) = \{\sigma_1,\ldots,\sigma_k\} \neq \emptyset, \sch{C}{C}(\sigma_i) = C_i\in \calC_T\\
    \lightning & \text{otherwise}
\end{cases}$;
    
    \item $\sch{C}{D \langle L \rangle (U)}(\sigma) = \begin{cases} \sch{D}{D}(\sigma)(\sch{L}{L}(\sigma), \sch{U}{U}(\sigma)) & \text{if defined} \\ \lightning & \text{otherwise}\end{cases}$.
\end{enumerate}

The semantics of component definitions is a map $\sch{D}{-} : {\rm Definitions} \longrightarrow (\calD \cup \{ \lightning \})^\Sigma,$
where Definitions is the set of all parse trees with root $D$.
For a scope $\sigma \in \Sigma$, we define $\sch{D}{-}(\sigma)$ as follows:

\begin{enumerate}
    \item  $\sch{D}{V}(\sigma) = \begin{cases}
    \sigma(\sch{V}{V}(\sigma)) & \text{if } \sch{V}{V}(\sigma) = x \in \dom(\sigma) \text{ and } \sigma(x) \in \calD \\
    \lightning & \text{otherwise}
\end{cases}$; 
    
    \item If $D$ is a component $\langle U_0 \rangle (U_1) \{C\}$, then for an array of parameter values $t \in \calV^\square$ and an array of nodes $q \in \calN^\square$, 
    we define $\sch{D}{\langle U_0 \rangle (U_1) \{C\}}(\sigma)(t,q)$ as follows:
    Recall from \cref{sec:syntax} that the number of parameters and nodes can implicitly define variables.
    Suppose that there exists a unique `index-defining' scope $\sigma' \in \Sigma$ such that for $m = \len(t)$ and $n = \len(q)$. Then we have
    \begin{enumerate}
    \item $\sch{U}{U_0}(\sigma') = [s_1,\ldots, s_m] \neq \lightning$ satisfies $s_i \simeq t(i)$, for all $1 \leq i \leq m$;
    \item $\sch{U}{U_1}(\sigma') = [p_1,\cdots, p_n] \neq \lightning$ satisfies $p_i \simeq q(i)$, for all $1 \leq i \leq n$;
    \item $\flatten([s_1,\ldots s_m,p_1,\ldots p_n]) \in \calN^\square$ has no duplicates;
    \item $\dom(\sigma') \subseteq \calN$ is minimal such that properties (a)-(c) are satisfied.
    \end{enumerate}
    We evaluate the body $C$ of the component definition to the component $\sch{C}{C}(\sigma\sigma')$, where $\sigma\sigma'$ is the composition of $\sigma$ and $\sigma'$.
    Define the map $r : \supp(\sch{C}{C}(\sigma\sigma')) \longrightarrow \calN$ as 
    \[ r(x) = \begin{cases}
            t_i(k_1)\cdots(k_l) & \text{if } x = s_i(k_1)\cdots(k_l) \\
            q_i(k_1)\cdots(k_l) & \text{if } x = p_i(k_1)\cdots(k_l) \\
            v \text{ fresh} & \text{otherwise} 
        \end{cases} \]
    Map $r$ is well-defined, because $\flatten([s_1,\ldots s_m,p_1,\ldots p_n]) \in \calN^\square$ has no duplicates.
    Note that $r$ is finite, since we assume that $\supp(\sch{C}{C}(\sigma\sigma'))$ is finite.
    We define $\sch{D}{\langle U_0 \rangle (U_1) \{C\}}(\sigma)(t,q)$ as the simultaneous substitutions
    $\sch{C}{C}(\sigma\sigma')[r(x)/x : x \in \dom(f)]$.
    If such `index-defining' scope $\sigma'$ does not exists or is not unique, then we simply define $\sch{D}{\langle U_0 \rangle (U_1) \{C\}}(\sigma)(t,q) = \lightning$.
\end{enumerate}

We define the semantics of files as a map $\sch{K}{-} : {\rm Files} \longrightarrow \Sigma \cup \{\lightning\},$
where Files is the set of parse trees with root $K$. 
Let $\sch{I}{-} : {\rm Imports} \longrightarrow \Sigma$ be the semantics of imports.
For a scope $\sigma \in \Sigma$, we define $\sch{K}{-}(\sigma)$ inductively as follows:
\begin{enumerate}
    \item $\sch{K}{I}(\sigma) = \sch{I}{I}$;
    \item $\sch{K}{KND}(\sigma) = \begin{cases} \sigma_0\{x \mapsto c\} & \text{if } \sigma_0 = \sch{K}{K}(\sigma) \neq \lightning, x = \sch{N}{N}\text{, and } c = \sch{D}{D}(\sigma_0) \neq \lightning\\ \lightning & \text{otherwise}\end{cases}$.
\end{enumerate}

\section{Input/output nodes}
\label{sec:direction}

As mentioned in \cref{sec:syntax}, nodes of primitive component definitions require input/output annotations.
Treo regards such port type annotations as attributes of the primitive component.
For a semantic sort $T$, we model the input nodes and output nodes of its instances via two maps $I,O : \calC_T \longrightarrow 2^\calN$ satisfying $\supp(C) = I(C) \cup O(C)$, for all $C \in \calC_T$.
If $x \in I(C) \cap O(C)$, then we call $x$ a {\em mixed} node.

\begin{figure}[t]
    \centering
    \subfigure[$P_1 \cdot P_2 \cdot P_3$]{
        \includegraphics[scale=1]{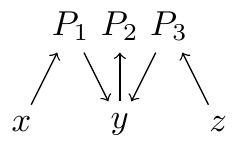}
        \label{fig:p1p2}
    }
    \qquad
    \subfigure[$\wfc(P_1 \cdot P_2 \cdot P_3)$]{
        \includegraphics[scale=1]{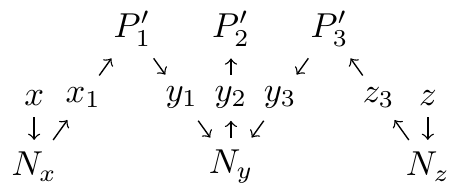}
        \label{fig:surgp1p2}
    }
    \caption{Surgery on an I/O-component to remove mixed nodes.}
    \label{fig:surg}
\end{figure}

\begin{example}[Mixed nodes]
    \label{ex:mixed}
    Recall the I/O component sort from \cref{ex:io}. Let $P_1 = (C_1, \{x\}, \{y\})$, $P_2 = (C_2, \{y\}, \emptyset)$, and $P_3 = (C_3, \{z\}, \{y\})$ be three primitive I/O components.
    \cref{fig:p1p2} shows a graphical representation of composition of $P_1$, $P_2$, and $P_3$.
    In this figure, an  arrow from a node $a$ to a component $P$ indicates that $a$ is an input node of $P$. 
    An arrow from a component $P$ to a node $a$ indicates that $a$ is an output node of $P$.    
    Node $y$ is an output node of $P_1$ and $P_3$, and it is an input node of $P_2$.
    Thus, $y$ is a mixed node in the composition $P_1 \cdot P_2 \cdot P_3$, where $\cdot$ is sequential composition of I/O components.
    \EOE
\end{example}

Most semantic sorts that distinguish input and output nodes assume {\em well-formed} compositions: 
each shared node in a composition is an output of one component and an input of the other.

\begin{definition}[Well-formedness]
    \label{defn:well-formed}
    A composition $C_1 \wedge_T \cdots \wedge_T C_n$, with $n \geq 0$, is well-formed if and only if 
    $|\{ i \in \{1,\ldots,n\} \mid x \in I(C_i)\}| \leq 1$ and $|\{i \in \{1,\ldots,n\} \mid x \in O(C_i)\}| \leq 1$, for all $x \in \calN$.
\end{definition}

For well-formed compositions, the behavior of the composition naturally corresponds to the composition of Reo connectors.
However, specification of complex components as well-formed compositions is quite cumbersome, because it requires explicit verbose expression of the `merge-replicate' behavior of every Reo node in terms of a suitable number of binary mergers and replicators. 
Reo nodes abstract from such detail and yield more concise specifications.  
Like Reo, Treo does not impose any restriction on the nodes of constituent components in a composition.
Indeed, the denotational semantics of components $\sch{C}{-}$ in \cref{sec:semantics} unconditionally computes the composition. % (even if the composition is not well-formed).
To define the semantics of $\sch{C}{-}$ for a semantic sort $T$ where $\wedge_T$ requires well-formedness, parsing a (non-well-formed) Treo composition needs the degree (i.e., the number of coincident input and output channel ends) of each node to correctly express the `merge-replicate' semantics of that node.
The degree of every node used in a definition can be known only at the end of that definition. 
The Treo compiler could discover the degree of every node via two-pass parsing.

Alternatively, Treo can delay applying composition $\wedge_T$ in $T$ until parsing completes, 
Treo accomplishes this by interpreting a Treo program over the I/O-component sort $\io_T$, as defined in \cref{ex:io}, wherein compositions consist of lists of primitive components.
First, Treo wraps each primitive component $C \in \calC_T$ within a primitive I/O-component $(C,I(C),O(C)) \in \calP_T$.
Using \cref{sec:semantics}, Treo parses the Treo program over the semantic sort $\io_T$ as usual, and obtains a single I/O-component $P_1 \cdots P_n \in \io_T$. 

However, the resulting composition  $P_1 \cdots P_n$ may not be well-formed.
Therefore, the Treo compiler applies some surgery on $P_1 \cdots P_n$ to ensure a well-formed composition.
This surgery consists of splitting all shared nodes in $X$, and reconnecting them by inserting a {\em node component}.
We model these node components (over semantic sort $T$) as a map $\node : (2^\calN)^2 \times \calN \longrightarrow \calC_T$.
For sets of names $I,O \subseteq \calN$ and a default name $x \in \calN$, the component $\node(I,O,x) \in \calC_T$ has input nodes $I$ (or $\{x\}$, if $I$ is empty) and output nodes $O$ (or $\{x\}$, if $O$ is empty).

\begin{definition}[Surgery]
    \label{defn:Surgery}
    The surgery map $\wfc : \io_T \longrightarrow \io_T$ is defined as
    $\wfc(P_1\cdots P_n) = P_1'\cdots P_n'\cdot \prod_{x \in \supp(P_1\cdots P_n)} N_x$,
    where $P_i' = P_i[x_i/x : x \in \supp(P_i)]$, for all $1 \leq i \leq n$, and
    $N_x = (\node(I_x,O_x,x),I_x,O_x)$, with
    $I_x = \{x_i \mid x \in O(P_i)\}$ and
    $O_x = \{x_i \mid x \in I(P_i)\}$. The composition $\prod$ is ordered arbitrarily.
\end{definition}

Intuitively, the surgery map takes a possibly non-well-formed composition and produces a well-formed composition by inserting node components. 
Although initially, multiple components may produce output at the same node.
After applying the surgery map, these components offer data for the same node component via different `ports'.

\begin{example}[Surgery]
    \label{ex:surgery}
    \cref{fig:surgp1p2} shows the result of applying the surgery map to the I/O-component $P_1 \cdot P_2 \cdot P_3$ from \cref{ex:mixed}.
    The surgery map consists of two parts.
    First, the surgery map splits every node $a \in \{x,y,z\}$ by renaming $a$ to $a_i$ in $P_i$, for every $1 \leq i \leq n$.
    Second, the surgery map inserts at every node $a \in \{x,y,z\}$ a node component $N_a$.
    Clearly, $\wfc(P_1\cdot P_2 \cdot P_3)$ is a well-formed composition.
    \EOE
\end{example}

\section{Related work}
\label{sec:related}

The Treo syntax offers a textual representation for the graphical Reo language \cite{DBLP:journals/mscs/Arbab04, DBLP:conf/birthday/Arbab11}.
We propose Treo as a syntax for Reo that (1) provides support for parameterization, recursion, iteration, and conditional construction; (2) implements basic design principles of Reo more closely than existing languages; and (3) reflects its declarative nature.
The graphical Reo editor implemented as an Eclipse plugin \cite{ECT} does not support parameterization, recursion, iteration, or conditional construction.
RSL (with CARML for primitives) \cite{DBLP:conf/coordination/BaierBKK09,DBLP:books/daglib/0031251} is imperative, while Reo is declarative.
FOCAML \cite{Jongmans16}, supports only constraint automata \cite{DBLP:journals/scp/BaierSAR06}, 
while Treo allows arbitrary user-defined semantic sorts for expressing the behavior of Reo primitives.

Since Treo leaves the syntax for primitive subsystems (i.e., semantic sorts) as user-defined, Treo is a ``meta-language'' that specifies compositional construction of complex structures (using the common core language defined in this paper) out of primitives defined in its arbitrary, user-defined sub-languages. 
As such, Treo is not directly comparable to any existing language.
We can, however, compare the component-based system composition of Treo with the system composition of an existing language. 

Treo components are similar to proctype declarations in Promela, the input language for the SPIN model checker developed by Holzmann \cite{DBLP:books/daglib/0020982}.
However, the focus of Promela is on imperative definitions of processes, while Treo is designed for declarative composition of processes.

SysML is a graphical language for specification of systems \cite{FMS14}. 
SysML offers 9 types of diagrams, including activity diagrams and block diagrams.
Each diagram provides a different view on the same system \cite{DBLP:journals/software/Kruchten95}.
Diagram types in SysML are comparable to semantic sorts in Treo.
The main difference between the two, however, is that Treo requires a well-defined composition operator, using which it allows construction of more complex components,
while diagram composition is much less prominent in SysML. 

A component model is a programming paradigm based on components and their composition.
Our Treo language can be viewed as one such component model with a concrete syntax. 
Over the past decades, many different component models have been proposed.
For example, CORBA \cite{OMG06} is a component model that is flat in the sense that every CORBA component is viewed as a black box, i.e., it does not support composite components. 
Fractal \cite{DBLP:journals/spe/BrunetonCLQS06} is an example of a component model that is hierarchical, which means a component can be a composition of subcomponents.
Concrete instances of Fractal consist of libraries (API's) for a variety of programming languages, such as Java, C, and OMG IDL \cite{DBLP:journals/spe/BrunetonCLQS06}.
Treo components and Fractal component differ with respect to interaction:
Treo components interact via shared names, while Fractal component interact via explicit bindings.

\section{Conclusion}
\label{sec:conclusion}

We propose Treo as a textual syntax for Reo connectors that allows user-defined semantic sorts, and incorporates Reo's predefined node behavior.
These features are not present in any of the existing alternative languages for Reo.
We provided an abstract syntax for Treo and its denotational semantics based on this abstract syntax.
We identify three possible directions for future work. 

First, since our semantics disallows recursion, a component in Treo is currently restricted to consist of a composition of finitely many subsystems.
Consequently, we cannot, for instance, express the construction of a primitive with an unbounded buffer, $B_\omega$, from a set of primitives with buffer capacity of one, $B_1$.
It seems, however, possible to use simulation and recursion to define $B_\omega$ in terms of $B_1$:
$B_\omega$ is the smallest (with respect to simulation) component that simulates $B_1$ and is stable under sequential composition with $B_1$.
These assumptions readily imply that $B_\omega$ simulates a primitive with buffer of arbitrary large capacity.
Semantically, the unbounded buffer would then be defined as a least fixed point of a certain operator on components.
An extension of Treo semantics that allows such fixed point definitions would provide a powerful tool to define complex `dynamic' components.

Second, the current semantics in \cref{sec:semantics} does not support components with an identity.
If we instantiate a component definition twice with the same parameters, we obtain two instances of the same component.
Ideally, component instantiation should return a component instance with a fresh identity.
Allowing components with identities in Treo enables programmers to design systems more realistically.

Finally, a semantic sort $T$ from \cref{defn:types} consists of a single composition operator $\wedge_T$.
Generally, a semantic sort consists of multiple composition operators (each with it own arity).
For example, we may need both sequential composition as well as parallel composition.
Extending Treo with (a variable number of) composition operators would enable users to model virtually all semantic sorts.

\bibliographystyle{eptcs} 
\bibliography{refsdblp,refsother}

\end{document}